\documentstyle[amssymb,12pt]{article}

\begin{document}

\title{Integrable Nonautonomous Nonlinear Schr{\" o}dinger Equations}
\author{ Metin G{\" u}rses\\
{\small Department of Mathematics, Faculty of Sciences}\\
{\small Bilkent University, 06533 Ankara - Turkey}}
\begin{titlepage}
\maketitle
\begin{abstract}
We show that a recently given nonautonomous nonlinear Schrodinger
equation (NLSE) can be transformed into the autonomous NLSE.
\end{abstract}
\end{titlepage}

\section{Introduction}

In \cite{gur} we have shown that integrable variable coefficient
KdV type of equations, $q_{t}=f(x,t)\,q_{3}+H(x,t,q,q_{1})$ are
transformable into autonomous integrable KdV type of equations,
$q_{t}=q_{3}+F(q,q_{1})$. It seems that this assertion is valid in
general. In a recent paper \cite{serkin} an integrable
nonautonomous NLSE is introduced

\begin{equation}\label{den0}
iQ_{t}+{D(t) \over 2}\, Q_{xx}+\sigma R(t)\,|Q|^2\,Q-[2
\alpha(t)\,x\,+{\Omega^2(t) \over 2}\,x^2\,]\,Q=0
\end{equation}

\noindent where the functions $R(t)$, $D(t)$, and $\Omega(t)$
representing the nonlinear, dispersion and the harmonic potential
terms respectively are restricted to satisfy the following
condition.

\begin{equation}
-\Omega^2(t)\, D(t)={d^2 \over dt^2} \ln D(t)+R(t) {d^2 \over
dt^2} {1 \over R(t)}-{d \over dt} \ln D(t)\, {d \over dt} \ln R(t)
\label{den1}
\end{equation}
They give a Lax pair of this equation where integrability implies
both (\ref{den0}) and (\ref{den1}). The spectral parameter is a
time dependent function and hence solitons obtained through the
inverse scattering method have time dependent amplitudes and
speeds. We shall first show that (\ref{den0}) is transformable to
the standard NLSE and produce the soliton solutions given in
\cite{serkin} directly from the soliton solution of the NLSE.

Let $Q=\Lambda\, q(X,T)$ where

\begin{equation}
T=G(t),~~~X=F(x,t) \label{eqn3},~~~\Lambda=\Lambda(x,t)
\end{equation}

\noindent Then Eq.(\ref{den0}) reduces to the standard NLSE

\begin{equation}
iq_{T}+q_{XX}+\sigma\,|q|^2\,q=0, \label{eqn1}
\end{equation}
where the following conditions must be satisfied.

\begin{eqnarray}
&&i\Lambda_{t}+{D(t) \over 2}\, \Lambda_{xx}-2 \alpha(t) x
\Lambda-{\Omega^2(t) \over 2} x^2 \Lambda=0, \\
&&i \Lambda F_{t}+{D(t) \over 2} [2 \Lambda_{x} F_{x}+\Lambda
F_{xx}]=0, \\
&&{D(t) \over 2}{F_{x}^2 \over {G_{t}}}=1,\,\,\,{ R |\Lambda|^2
\over {G_{t}}}=1
\end{eqnarray}

One can find a solution of the above equations. One solution is
given as follows

\begin{eqnarray}
&&\Lambda=r(t)e^{i\theta}, ~~\theta=-{{r_{t}} \over D(t)
r}\,x^2+\alpha_{1}\,x+\alpha_{2},\\
&&r^2=2 r_{0}^4   {R(t) \over D(t)}, \label{den2}
\end{eqnarray}
where $\alpha_{1}$ and $\alpha_{2}$ are functions of $t$, $r_{0}$
is an arbitrary real constant and $r$ is subjected to satisfy the
equation

\begin{equation}
{d \over dt}\left({1 \over r(t)D(t)} {dr \over dt}\right)-{D(t)
\over 2} \left({2 \over r(t)D(t)} {dr \over dt}
\right)^2-{\Omega^2(t) \over 2}=0 \label{den3}
\end{equation}
When $r(t)$ given in (\ref{den2}) is used in (\ref{den3}) we
obtain exactly the condition in (\ref{den1}). Furthermore we
obtain $F={r^2 \over r_{0}^2}\,x+F_{1}$ and

\begin{eqnarray}
&&\alpha_{1}=-2 {R(t) \over D(t)} \, \int {\alpha(t) D(t) \over
R(t)}dt,\\
&&\alpha_{2}=-{1 \over 2}\, \int\, D(t)\, \alpha_{1}\,^2 dt,\\
&& F_{1}=-r_{0}^2\, \int \, \alpha_{1}(t)\,R(t) dt,\\
&&G(t)=2r_{0}^4 \int\,{R^2(t) \over D(t)}\,dt
\end{eqnarray}

One soliton solution of the NLSE (\ref{eqn1}) was  given by
Zakharov and Shabat \cite{zak} which depends on the sign
$\sigma$. For $\sigma=1$ we have

\begin{equation}
q(X,T)= a \exp \left[ i\left \{ {c \over 2}(X-cT)+nT \right \}
\right]\, \mbox{sech}\left\{a(X-cT)/\sqrt{2}\right \}
\end{equation}
where $4n=c^2+2a^2$. For $\sigma=-1$ we have

\begin{equation}
q(X,T)={1 \over \sqrt{2}}\, [c-2i\kappa \tanh\kappa(X-cT)]e^{-imt}
\end{equation}
where $c^2=2m -4 \kappa^2$. Inserting the transformations
(\ref{eqn3}) in the above Zakharov-Shabat solutions we obtain the
solutions obtained by Serkin et al \cite{serkin}.

I would like to thank Atalay Karasu for discussions.  This work is
partially supported by the Turkish Academy of Sciences.

\end{document}